\documentclass[fleqn,usenatbib]{mnras}
\usepackage{amsmath}
\usepackage{graphicx}  % needed for figures
\usepackage{dcolumn}   % needed for some tables
\usepackage{bm}        % for math
\usepackage{verbatim}   % for math
\usepackage{multirow}
\usepackage{ulem}
\usepackage{xcolor}

\usepackage{hyperref}
\hypersetup{colorlinks=true}

\usepackage{newtxtext,newtxmath}
\usepackage[T1]{fontenc}
\usepackage{ae,aecompl}

\usepackage{lineno}
\pubyear{0000}

\title{Self-calibration method for II and GI types of intrinsic alignments of galaxies}

\author[Ji Yao et al.]{
	Ji Yao,$^{1,2}$\thanks{E-mail: Ji.Yao@outlook.com}
	Mustapha Ishak,$^{1}$\thanks{E-mail:mishak@utdallas.edu}
	M. A. Troxel$^{3}$\thanks{E-mail:michael.a.troxel@gmail.com}
    \newauthor
    (The LSST Dark Energy Science Collaboration)
	\\
    $^{1}$Department of Physics, The University of Texas at Dallas, Dallas, TX 75080, USA\\
    $^{2}$Department of Astronomy, Shanghai Jiao-Tong University, Shanghai 200240, China\\
	$^{3}$Department of Physics, Ohio State University, Columbus, OH 43210, USA\\
}

\date{Accepted XXX. Received YYY; in original form ZZZ}

\begin{document}
	
	\label{firstpage}
	\pagerange{\pageref{firstpage}--\pageref{lastpage}}
	\maketitle
	
    \begin{abstract}
    We introduce a self-calibration method that can be applied to the intrinsic ellipticity--intrinsic ellipticity (II) and gravitational shear -- intrinsic ellipticity (GI) types of intrinsic alignment of galaxies. The method combines previous self-calibration techniques with modifications to one of them in order to use auto-spectra in addition to cross-spectra between redshift bins. This allows one to use the self-calibration while preserving all the constraining power of cosmic shear from surveys. We show that the new method provides more flexibility in using various redshift bin widths. We perform cosmological parameter constraint forecast when this method is applied to the Large Synoptic Survey Telescope (LSST). Compared to the original self-calibration, we find that the new method provides further significant reduction in any residual shift in the cosmological parameters (e.g. factors of $2-4$ for the dark energy equation of state) which is promising for accurate cosmology. 
    
    	\end{abstract}
    
	\section{Introduction}
	
	With highly precise surveys and data, systematic effects have moved at the forefront of interest in cosmological analyses. This is particularly the case for cosmic shear which promises to put stringent constraints on the properties of dark energy and to test gravity at cosmological scales \citep{Kaiser1992,Hu1999,Heavens2000,Bacon2001,Ishak2005,Ishak2007,Joudaki2009,Weinberg2013,Ishak2006,Linder2007,Heavens2009,Dossett2012,Dossett2013}

	Intrinsic alignments (IA) of galaxies constitute one of the most serious systematic effects that contaminate cosmic shear, see for example  \cite{Bacon2001,Bernstein2002,Erben2001,Heymans2004,Hirata2003,Ishak2004,BridleKing,Faltenbacher2009}. 
	There has been a continuous effort in the community to understand, model and mitigate various types of IA.
	These correlations exist for 2-point and 3-point correlations and affect both the shear spectrum and bi-spectrum \citep{Bernstein2009,Troxel2012,Joachimi2013,Krause2016,Blazek2015}.
	The intrinsic-intrinsic shear correlation noted as (II) and (III) for the 3-point are due to alignment of galaxies resulting from their formation in the same tidal gravitational field. Other mixed correlations, such as the gravitational - intrinsic shear (GI) or the 3-points (GGI, GII), come from combinations of foreground galaxies  aligned with dark matter structures that are also correlated with tangentially sheared images of background galaxies \citep{Catelan2001,Hirata2003,King2005,Mandelbaum2006,Hirata2007,Semboloni2009,Troxel2012,Refregier2003}. We refer the reader to some recent reviews and developments on the topic of intrinsic alignments \citep{TroxelIshak,Kiessling2015,Kirk2015,Joachimi2015,Kilbinger2015,Blazek2017,Troxel2017}.
	
	At least two routes have been taken by the astrophysics community in order to mitigate the IA systematics. 
	One route is to use existing models of IA, add their nuisance parameters to the list of cosmological parameters, and implement them in the analysis pipelines. Then, the survey data is used to simultaneously constrain all the parameters, including those of the IA systematics. This is commonly referred to as the marginalization method \citep{Krause2016,Joachimi2013,Joachimi2015}. With incoming and future precision experiments, the method will require sophisticated IA models, and some progress is being made in this direction \citep{BridleKing,Krause2016,Blazek2015,Joachimi2013,Joachimi2015}. \textcolor{black}{Modeling IA according to different types of the galaxies, for example red/blue galaxies or elliptical/spiral galaxies, are also discussed in recent works \citep{Blazek2017,Tugendhat2017}.}
	The second route is to use methods that aim at measuring or isolating the IA signal(s) and reducing them, or subtracting them, from the cosmic shear data.
	For example,  for the II and III, it is often proposed to use only the cross-spectra of galaxies with bins sufficiently far so the II and III signals die away since it is present between relatively close galaxies \citep{Schneider2010,SC2008}. However, this does not work for the GI/GII/GGI types of correlations since they exist even between remote galaxies as well. Another technique for this second route is to throw away some samples of galaxies where a particular type of correlation is known to be prominent, such as the rejection of Luminous Red Galaxies in some samples to get rid of the GI, e.g. \cite{Okumura2009,Dossett2013,Blazek2015,Chisari2017}. Also, nulling techniques have been proposed in \cite{Huterer2005,Joachimi2008,Joachimi2009} where some of the lensing signals are also sacrificed as well. Finally, self-calibration methods for the 2-point \citep{SC2008,Zhang2010} and 3-point \citep{Troxel2012,Troxel2012b,Troxel2012c,TroxelIshak} have been proposed to mitigate with the GI, GGI and GII type of signals. The idea is to derive some scaling relations that will allow one to determine the GI (GGI, GII) signals from the same cosmic shear survey observables and then subtract them out, as we describe further in the paper. 
	
	In the original paper on the self-calibration \citep{SC2008}, Zhang showed that the SC can clean the signal at the 90\% level even if SC works at its lowest efficiency. This was confirmed in \cite{Troxel2012}, and similar efficiency was also found for the 3-point functions. The SC method is an important check and alternative method to explore IA, as it can not only clean the IA signal, but can potentially measure the signal and make it available for IA model studies. It is worth clarifying that SC methods were not meant to necessarily maximize cosmological information, but to give an independent measurement of the IA correlations.
    
	In a previous paper, \cite{Yao2017} investigated the effects of applying self-calibration to future photo-z surveys such as LSST, WFIRST and Euclid. We evaluated the loss of constraining power when the SC is applied to the GI alignments and found it to be comparable  to that of the marginalization method. We also found significant improvement in the accuracy of the best-fit cosmological parameters, \textcolor{black}{compared with the case with full IA contamination}. However, it was pointed out there that the accuracy of the best-fit cosmological parameters and the gain from the SC can be further improved if one could find a way to mitigate the II signal within the same and neighboring redshift bins. 
    
	In this work, we present an improved SC method where the technique deals not only with the GI signal, but also the II signal providing more constraining power in a combined method that significantly improves the accuracy of the cosmological parameter constraints. The new method allows the self-calibration to use the auto-spectra in addition to the cross-spectra between redshift bins which maximizes the constraining power of cosmic shear from surveys. We show that the new SC is complementary to the original one and the two methods can be used jointly. The major part of IA signal is still measured by SC2008 \citep{SC2008}, while the additional method uses a modified version of SC2010 \citep{Zhang2010}, which is recently verified in a study with N-body simulation \cite{Meng2018}.
    
	The paper is organized as follows. In Section \ref{subsection SC2008} we give an overview of the 2-point Self Calibration technique (SC2008) of IA, with a summary of the forecast done  in our previous work. We briefly review the SC2010 method and add some  modifications to it in Section \ref{s:SC 2010 comment}. Next, by combining the SC2008 method introduced in Section \ref{subsection SC2008} and the modified SC2010 method introduced in Section \ref{s:SC 2010 comment}, we present a new SC method SC2017 that we apply to the IA spectra in Section \ref{s:SC II} and the GI spectra in \ref{s:SC GI}. The extra measurement error introduced by the SC2017 method is discussed in Section \ref{s: error}. The survey specifications and theoretical models we use for IA, photo-z, galaxy bias and shift in cosmological parameters are included in Section \ref{section models}. We show results of the improvement obtained from using SC2017 by performing cosmological parameters constraint forecasts for an LSST-like survey in Section \ref{section results}. A summary is provided in Section \ref{section summary}.
	
	%%%%%%%%%%%%%%%%%%%%%%%%%%%%%%%%%%%%%%%%%%%%%%%%%%%%%%%%%%%%%%%%%%%%%%%
	%%%%%%%%%%%%%%%%%%%%%%%%%%%%%%%%%%%%%%%%%%%%%%%%%%%%%%%%%%%%%%%%%%%%%%%%
	\section{Overview of the Original Self-Calibration (SC) Methods} \label{section SC2008}
	
	%%%%%%%%%%%%%%%%%%%%%%%%%%%%%%%%%%%
	\subsection{The 2008 SC method (SC2008)} \label{subsection SC2008}
	%%%%%%%%%%%%%%%%%%%%%%%%%%%%%%%%%%%
	
	The measured shear contains $\gamma^G+\gamma^I+\gamma^N$, where the superscript G stands for gravitational shear, I for intrinsic alignment, and N for shot noise, including the measurement noise and the random noise. The observed angular cross-correlation power spectra are contaminated by the intrinsic alignment components. The power spectra are given by \citep{BridleKing}:
	\begin{eqnarray}
	\label{observables}
	\textit{C}^{(1)}_{ij}(l)&=&\textit{C}^{GG}_{ij}(l)+\textit{C}^{IG}_{ij}(l)+\textit{C}^{GI}_{ij}(l)+\textit{C}^{II}_{ij}(l)\notag\\
    &+&\delta_{ij}\textit{C}^{GG,N}_{ii},\\
	\textit{C}^{(2)}_{ii}(l)&=&\textit{C}^{gG}_{ii}(l)+\textit{C}^{gI}_{ii}(l),\\
	\textit{C}^{(3)}_{ii}(l)&=&\textit{C}^{gg}_{ii}(l)+\delta_{ij}\textit{C}^{gg,N}_{ii},
	\end{eqnarray}
where the three equations represent respectively the observed shape-shape, galaxy-shape and galaxy-galaxy power spectra. 
	
	In Eq.\,(\ref{observables}), the shear spectra $\textit{C}^{GG}$ is what is used in Weak Lensing (WL) to constrain the cosmological model. A Self-Calibration (SC) technique was introduced by Zhang in 2008 in order to subtract the major contamination term, $\textit{C}^{IG}_{ij}$, as well as to minimize the effect of $\textit{C}^{GI}_{ij}$ and $\textit{C}^{II}_{ij}$ \citep{SC2008}. It was shown there that, under the small-bin condition (i.e. with tomographic bin-size $\Delta z^P\sim0.2$ where $z^P$ stands for photometric redshift), the parts of $\textit{C}^{GI}_{ij}$ and $\textit{C}^{II}_{ij}$ for $i<	j$ are negligible \textcolor{black}{due to the fact that they are local effects}, while the contamination term $\textit{C}^{IG}_{ij}$ can be determined by a scaling relation without any a prior knowledge about the intrinsic alignment model. The scaling relation reads 
	\begin{equation} \label{scaling-1}
	\textit{C}^{IG}_{ij}(l)\simeq \frac{W_{ij}\Delta_i}{b_i(l)}\textit{C}^{Ig}_{ii}(l),
	\end{equation}
	where 
	\begin{subequations}
		\begin{align} 
		W_{ij}&\equiv \int_{0}^{\infty}dz_L\int_{0}^{\infty}dz_S
		[W_L(z_L,z_S)n_i(z_L)n_j(z_S)],\label{Wij} \\ 
		\Delta_i^{-1}&\equiv \int_{0}^{\infty}n_i^2(z)\frac{dz}{d\chi}dz. \label{Delta_i}
		\end{align}
	\end{subequations}
	Here $i<j$ is required, as the approximation of Eq.\,\eqref{scaling-1} has a low efficiency in the auto-spectra ($i=j$), which is an important feature that will be discussed in the next subsection. Of course, the other limitation is that the SC2008 method cannot be applied to the $\textit{C}^{II}_{ij}$ term. 
	
	In Eq.\,\eqref{Wij}, $W_L$ is the lensing kernel:
	\begin{equation}
	W_L(z_L,z_S)=\begin{cases}
	\frac{3}{2}\Omega_m\frac{H_0^2}{c^2}(1+z_L)\chi_L(1-\frac{\chi_L}{\chi_S}) &\text{for $z_L<z_S$}\\ 0 &\text{otherwise}
	\end{cases},
	\end{equation}

	In Eq.~(\ref{scaling-1}), $b_i(l)\approx\int_0^\infty b_g(k,z)n_i(z)dz$ is the averaged galaxy bias in each redshift bin, where $b_g$ is the galaxy bias given by $b_g(k=l/\chi,z)=\delta_g/\delta$ (this approximation is used in this theoretical forecasting work while in practice, the averaged bias, $b_i$, is calculated using another approximation $\textit{C}^{gg}\approx b_i^2\textit{C}^{mm}$, where $\textit{C}^{mm}$ is the matter power spectrum which can be inferred from CMB experiments.) Finally, $n_i(z)$ is the normalized true-z distribution of the i-th tomographic bin. 
	
	As shown in \cite{SC2008}, the measurement of ${C}^{Ig}_{ii}$ in Eq.\,\eqref{scaling-1} is given by another relation based on the difference between I-g and G-g correlation as follows:
	\begin{equation} \label{scaling-2}
	\textit{C}^{Ig}_{ii}(l)=\frac{\textit{C}^{(2)}_{ii}|_S(l)-Q_i(l)\textit{C}^{(2)}_{ii}(l)}{1-Q_i(l)},
	\end{equation}
    and where a new observable, $\textit{C}^{(2)}_{ii}|_S(l)$, is introduced. The subscript ``S'' stands for correlating pairs for which $z^P_G<z^P_g$ is true ($z^P_G$ stands for the photo-z of the object with the shear signal used in the correlation, while $z^P_g$ is the photo-z of the object corresponding to the galaxy number count used). $Q_i$ is calculated as
	\begin{equation} \label{Q}
	Q_i(l)\equiv \frac{\textit{C}^{Gg}_{ii}|_S(l)}{\textit{C}^{Gg}_{ii}(l)}
	\end{equation}
and depends on the quality of the survey's photo-zs. When considering the data, the value of $Q_i$ can be estimated either from Eq.\,\eqref{Q} by applying the observed redshift distribution to the spectra above, or using an estimator integrated from the redshift distribution, as discussed in \cite{SC2008}. We use here the latter.

The derivation of Eq.\,\eqref{scaling-1} and \eqref{scaling-2}, the impact of photo-z errors and the uncertainty in $Q_i$ are all well-discussed in \cite{SC2008}. Using these two equations, one can estimate the IG signal from the survey. The efficiency and accuracy of the SC2008 method was verified and confirmed in \cite{Troxel2012}, where the authors also expanded the method to the 3-point correlations and bi-spectra. 
	
	In our previous forecast work using SC2008, we derived a Fisher matrix formalism for the SC method and applied it to obtain confidence contours for cosmological parameters before and after applying the SC. We also employed a Newton-based method to calculate the residual shift of cosmological parameters after applying SC2008. The shift in the best-fit cosmological parameters using SC2008 was shown to be under the $1\sigma$ level, so that the majority of the IA contamination had been indeed removed from cosmic shear signal. 
    
    We also showed there that the covariance of the measured GG spectra using SC2008 can be approximated by only using the covariance of the theoretical cosmic shear spectra. That is 
	\begin{align} \label{cov approx}
	Cov(\textit{C}^{GG(SC)}_{ij}(\ell),\textit{C}^{GG(SC)}_{pq}(\ell))
	\approx Cov(\textit{C}^{GG}_{ij}(\ell),\textit{C}^{GG}_{pq}(\ell)).
	\end{align}
	This is in agreement with the estimates from \citep{SC2008} that the extra error introduced by SC from the measurements of $C^{Ig}$, $b_i$, and $Q_i$ is only of $\sim10\%$, at most. 
	
	Despite its competitive performance and efficiency when calculating forecasts for LSST, WFIRST and Euclid (see for example \cite{Yao2017}), there is room for further improvement for the SC2008.
    
	First, SC2008 does not apply to auto-spectra ($i=j$) since it has low accuracy ($\sim 80\%$ level) in measuring $\textit{C}^{IG}$. Second, SC2008 cannot clean the non-negligible $\textit{C}^{II}$ in the auto-spectra. As we explained in the previous subsection, the binning method can only minimize the effects of $\textit{C}^{GI}_{ij}$ and $\textit{C}^{II}_{ij}$ so that they are less dominant for $i\ne j$, but when $i=j$ they are at the same level as $\textit{C}^{IG}_{ij}$  and thus are not negligible. So, when applying SC2008, we cannot use the auto-spectra which results in a loss in constraining power. Similarly, in the adjacent bins ($i+1=j$), the efficiency is not ideal ($\sim 90\%$ in $\textit{C}^{IG}$ and $\textit{C}^{II}$ is not negligible). In our previous paper, we showed how these aspects can affect the forecast with larger confidence contours and larger shifts in the best-fit cosmological parameters. For future surveys with better quality photo-zs and redshift bin sizes less than $0.2$, these limitations will become more pronounced and require remedy. 

	%%%%%%%%%%%%%%%%%%%%%%%%%%%%%%%%%%%%%%%%%%%%%%%%%%%%%%
	\subsection{ The 2010 SC Method (SC2010) plus a Modification}
	\label{s:SC 2010 comment}
	%%%%%%%%%%%%%%%%%%%%%%%%%%%%%%%%%%%%%%%%%%%%%%%%%%%%%%%%
	
	In the 2010 SC paper \cite{Zhang2010}, the author found that when correlating two redshift bins, the dependencies of the intrinsic spectra and the galaxy spectra on the redshift are very close, so that the following approximations apply:
	%\begin{widetext}
	\begin{eqnarray}
	\label{scaling-II}
	&\textit{C}^{II}(\Delta z^P|\ell,\bar{z}^P)\approx A_{II}(\ell,\bar{z}^P)\textit{C}^{gg}(\Delta z^P|\ell,\bar{z}^P)~, \\
	\label{scaling-Ig}
	&\textit{C}^{Ig}(\Delta z^P|\ell,\bar{z}^P)\approx A_{Ig}(\ell,\bar{z}^P)\textit{C}^{gg}(\Delta z^P|\ell,\bar{z}^P)~, \\
	&\textit{C}^{GI}(\Delta z^P|\ell,\bar{z}^P)+\textit{C}^{IG}(\Delta z^P|\ell,\bar{z}^P) \approx A_{GI}(\ell,\bar{z}^P) \notag \\
	\label{scaling-GI}
	&\times \left[ \textit{C}^{Gg}(\Delta z^P|\ell,\bar{z}^P)+\textit{C}^{gG}(\Delta z^P|\ell,\bar{z}^P) \right]~,
	\end{eqnarray}
	%\end{widetext}
	where for two photo-z bins with mean photo-z $z_1^P$  and $z_2^P$, the mean redshift $\bar{z}^P\equiv (z_1^P+z_2^P)/2$ and redshift separation $\Delta z^P\equiv z_2^P-z_1^P$ have been defined. The above approximations \,\eqref{scaling-II}, \eqref{scaling-Ig} and \eqref{scaling-GI}, were derived from studying the redshift evolution of the related spectra  \,\citep{Zhang2010}. The author found that when changing the redshift bin separation, the gg, Ig and II spectra evolve in a similar way, while the Gg+gG and GI+IG spectra evolve differently than the gg, Ig and II spectra. The author also tested these three approximations and found them to work very well. 
    
    In his method, $A_{II}$, $A_{Ig}$ and $A_{GI}$ are treated as free parameters, while the ratio $C^{II}/C^{gg}$, $C^{Ig}/C^{gg}$ and $(C^{GI}+C^{IG})/(C^{Gg}+C^{gG})$ remain almost constant. Eq.\,\eqref{scaling-II} and Eq.\,\eqref{scaling-Ig} were tested to be accurate at the $\sim 1\%$ level. Eq.\,\eqref{scaling-GI} is not as good, but still accurate at the $\sim 10\%$ level.
	
	We observe here that the approximations above can also be obtained from the small-bin approximation. Indeed, the Limber integrals of II, Ig and gg spectra read 
	\begin{subequations}
		\begin{align} 
		\textit{C}^{II}_{ij}(\ell)&=\int_{0}^{\infty}\frac{n_i(\chi)n_j(\chi)}{\chi^2} P_{II}\left(k=\frac{\ell}{\chi};\chi\right) d\chi \label{II}, \\
		\textit{C}^{Ig}_{ij}(\ell)&=\int_{0}^{\infty}\frac{n_i(\chi)n_j(\chi)}{\chi^2}b_g(k,\chi) P_{I\delta}\left(k=\frac{\ell}{\chi};\chi\right) d\chi \label{Ig}, \\
		\textit{C}^{gg}_{ij}(\ell)&=\int_{0}^{\infty}\frac{n_i(\chi)n_j(\chi)}{\chi^2}b_g(k,\chi)^2 P_\delta\left(k=\frac{\ell}{\chi};\chi\right) d\chi \label{gg}
		\end{align}
	\end{subequations}
%	where $M_{\rm IA}$ stands for the model that creates the underlying IA signals, $b_g$ the galaxy bias and $n_i(\chi)$ the galaxy distribution in the i-th photo-z bin. The only difference is whether it is the $M_{\rm IA}$ or the $b_g$ within the integral. By applying the small binning method, which separates photo-z into bins with small widths (for example $\Delta z^P=0.2$ as used in the previous Ref.\,\citep{SC2008}, \citep{Zhang2010} and our previous work \citep{Yao2017}), the following approximations are obtained by connecting Eq.\,\eqref{II} and \eqref{gg}, or by connecting Eq.\,\eqref{Ig} and \eqref{gg}:
	where $P_{II}$, $P_{I\delta}$ are the 3-D IA power spectra, $b_g$ is the galaxy bias and $n_i(\chi)$ is the galaxy distribution in the i-th photo-z bin. 
    
    Now, by applying the small binning method, which separates photo-z into bins with small widths (for example $\Delta z^P=0.2$ as used in \cite{SC2008}, \cite{Zhang2010} and \cite{Yao2017}), the following approximations are obtained by combining Eq.\,\eqref{II} and \eqref{gg}, or by combining Eq.\,\eqref{Ig} and \eqref{gg}:
	\begin{subequations}
		\begin{align} 
		\label{my-II}
		\textit{C}^{II}_{ij}(\ell)&\approx A^{II}_{ij}(\ell)\textit{C}^{gg}_{ij}(\ell)~, \\
		\label{my-Ig}
		\textit{C}^{Ig}_{ij}(\ell)&\approx A^{Ig}_{ij}(\ell)\textit{C}^{gg}_{ij}(\ell)~.
		\end{align}
	\end{subequations}
	Here $A^{Ig}_{ij}$ results form transferring one ``g'' (galaxy) signal into one ``I'' (intrinsic alignment) signal in the cross-correlation of bin-pairs \{i,j\}. Similarly, $A^{II}_{ij}$ results from transferring both ``g'' signals into ``I'' signals in \{i,j\} pairs.

	It is worth noting that in Zhang's 2010 SC paper, the variables ``$A$'' in Eqs.\,\eqref{scaling-II}, \eqref{scaling-Ig} and \eqref{scaling-GI} are treated as free parameters. As such, these equations can be viewed as rather another modeling method for IA. Additionally, because the amplitudes, $A^{II}$ and $A^{Ig}$, are outside the Limber integration, such a modeling method is less appealing than other advanced IA modeling where the model parameters are inside the Limber integrand and some physical motivations are usually taken into consideration. Therefore, instead of using SC2010 as proposed, we introduce an amended (new) SC2017 method in the following section.
	
 We calculate the various spectra used in this work using CosmoSIS software \citep{CosmoSIS}. We use the IA, photo-z and galaxy bias models as defined further in Section \ref{section models}. Similarly to \cite{Yao2017}, we choose the angular scale range as $20\leq\ell\leq5000$.
	
	%%%%%%%%%%%%%%%%%%%%%%%%%%%%%%%%%%%%%%%%%%%%%%%%%%%%%%%%%%%%%%%%
	%%%%%%%%%%%%%%%%%%%%%%%%%%%%%%%%%%%%%%%%%%%%%%%%%%%%%%%%%%%%%%%%
	\section{A new self-calibration of II and GI types of IA}
	\label{s:SC2017}
	%%%%%%%%%%%%%%%%%%%%%%%%%%%%%%%%%%%%%%%%%%%%%%%%%%%%%%%%%%%%%%%%
	
	As discussed in the previous section, it is preferable not to model the SC ``transfer amplitudes'', $A^{II}$ and $A^{Ig}$. Therefore, we attempt to find a way to rather directly measure IA including both II and GI. We show that this can be accomplished by some specific successive steps where the small bin approximation plays an important role. The bin-width in this work is chosen to be $\Delta z^P=0.1$, which represents an improvement compared with $\Delta z^P=0.2$ used in SC2008. For comparison with \cite{SC2008} and \cite{Yao2017}, we have also tested $\Delta z^P=0.2$ and found that the efficiency of SC2017 is at a similar level as that using $\Delta z^P=0.1$. More details about the photo-z, IA and bias model used in this work are provided in Section \ref{section models}.

	%%%%%%%%%%%%%%%%%%%%%%%%%%%%%%%%%%%%%%%%%%%%%%%%%%%%%%%%%%%%
	\subsection{Application to the II-type of IA}
	\label{s:SC II}
	%%%%%%%%%%%%%%%%%%%%%%%%%%%%%%%%%%%%%%%%%%%%%%%%%%%%%%%%%%%%
	
	\begin{figure}
		\includegraphics[width=1.0\columnwidth]{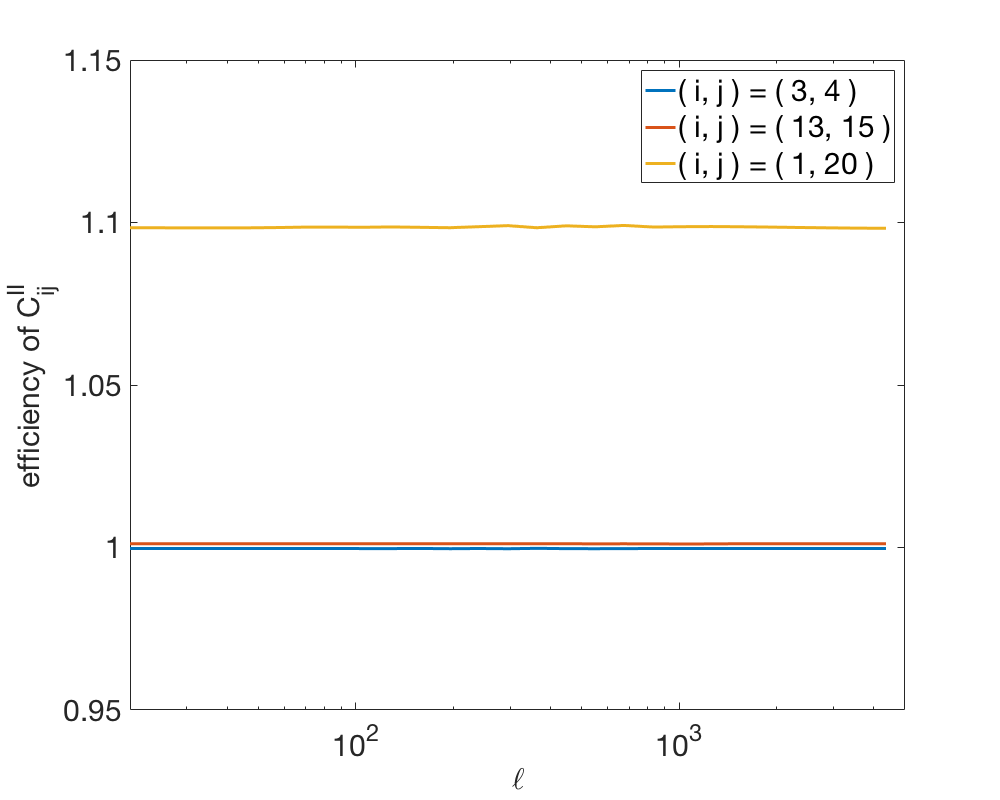}
		\caption{II efficiency with different bin-pairs. Y-axis represents the efficiency, defined by $\textit{C}^{II(SC)}/\textit{C}^{II({\rm true})}$. X-axis represents the angular modes $\ell$. For all the bin-pairs and the total range of $\ell$, the accuracy is very close to 1. The largest difference appears when the two bins are farthest apart ($i=1$, $j=20$), but the difference is only at $\sim 10\%$. However, for these cases where the efficiency deviates significantly from 1, the II signal itself is already negligible.
			\label{fig II efficiency}}
	\end{figure}
	
	First we take into consideration that using the small bin approximation and changing one ``g'' to one ``I'' in the cross power spectrum requires multiplication by $A^{Ig}$, see for example Eq.\,\eqref{my-Ig}. It follows:
	\begin{equation}
	A^{II}_{ij}(\ell)\approx \left(A^{Ig}_{ij}(\ell)\right)^2\,. \label{Ig->II}
	\end{equation}
	
	We numerically tested the above approximation by calculating $A^{II}_{ij}/(A^{Ig}_{ij})^2$, where the parameter ``A'' is inferred from Eq.\,\eqref{my-II} and \eqref{my-Ig}. The result found is very close to 1, meaning that the approximation of Eq.\,\eqref{Ig->II} works well for all bin-pairs and all $\ell$ values used. This approximation is accurate with the difference at $\sim 0.1\%$ level. In the numerical calculations, we used the linear IA model and linear galaxy bias model. For the non-linear models, the accuracy of this approximation will depend on the correlation between the IA model and the bias model and there will be some loss of efficiency, see the Appendix.
	Thus, with Eq.\,\eqref{Ig->II},	 the unknown parameters $A^{II}$ and $A^{Ig}$ reduce to one, $A^{Ig}$. However, the estimation of $A^{Ig}_{ij}$ is still required. Again, when small bin approximation is applied, another relation can be derived as follows:
	\begin{equation}
	A^{Ig}_{ij}(\ell)\approx \sqrt{A^{Ig}_{ii}(\ell)A^{Ig}_{jj}(\ell)} \label{ii->ij}.
	\end{equation}
	We tested this approximation as done for the previous one and found it to be good at the level of $0.1\%$ to $1\%$ for the bin-pairs and $\ell$ values used. Now, the measurement of $A^{Ig}_{ii}$ will directly lead to an estimation of $\textit{C}^{Ig}_{ij}$ and $\textit{C}^{II}_{ij}$ through Eq.\,\eqref{my-II} and \eqref{my-Ig}.
	
	The estimation of $A^{Ig}_{ii}$ comes from combining Eq.\,\eqref{my-Ig} and Eq.\,\eqref{scaling-2}. Through Eq.\,\eqref{scaling-2}, the information about IA can directly be measured as $\textit{C}^{Ig}_{ii}(\ell)$ by using the information in photo-zs. Thus by following the steps below, the information on IA is obtained as follows:
	\begin{equation} \label{steps1}
	\textit{C}^{Ig}_{ii} \stackrel{Eq.\,\eqref{my-Ig}}{\longrightarrow} A^{Ig}_{ii} \stackrel{Eq.\,\eqref{ii->ij}}{\longrightarrow} A^{Ig}_{ij} \stackrel{Eq.\,\eqref{Ig->II}}{\longrightarrow} A^{II}_{ij} \stackrel{Eq.\,\eqref{my-II}}{\longrightarrow} \textit{C}^{II}_{ij}\,.
	\end{equation}
	Each step is an approximation but with a tested and well accepted level of accuracy. As shown in Fig.\,\ref{fig II efficiency}, the final accuracy of $\textit{C}^{II}$ is very good and generally within $\sim 1\%$ of the true (simulated data) II spectra for most bin-pairs. The approximation is no longer precise only when considering very large bin separation; however, in this case, the II signals are already very small and negligible.
	
	The redshift distribution in this work is chosen to be for a LSST-like survey, see Table.\,\ref{tab surveys} and Fig.\,\ref{fig n(z)}. Fig.\,\ref{fig II efficiency} only shows 3 bin-pairs for readability but shows the cases of most interest. The bin pair (3,4) shows the efficiency of adjacent bins at low-z, while (13,15) shows the efficiency of close bins at high-z. The II signal is not negligible at close bins like these, where our SC2017 of II works extremely well. The (1,20) pair shows the efficiency at largest redshift separation in this work, where the SC2017 is at its worst efficiency, but the II signal itself is extremely small and negligible in this case. Thus, Fig.\,\ref{fig II efficiency} confirms  the expected accuracy of SC2017 for the II term.
	
	%%%%%%%%%%%%%%%%%%%%%%%%%%%%%%%%%%%%%%%%%%%%%%%%%%%%%%%%%%%%
	\subsection{Application to the GI-type of IA}
	\label{s:SC GI}
	%%%%%%%%%%%%%%%%%%%%%%%%%%%%%%%%%%%%%%%%%%%%%%%%%%%%%%%%%%%%
	
\begin{figure}
		\includegraphics[width=1.1\columnwidth]{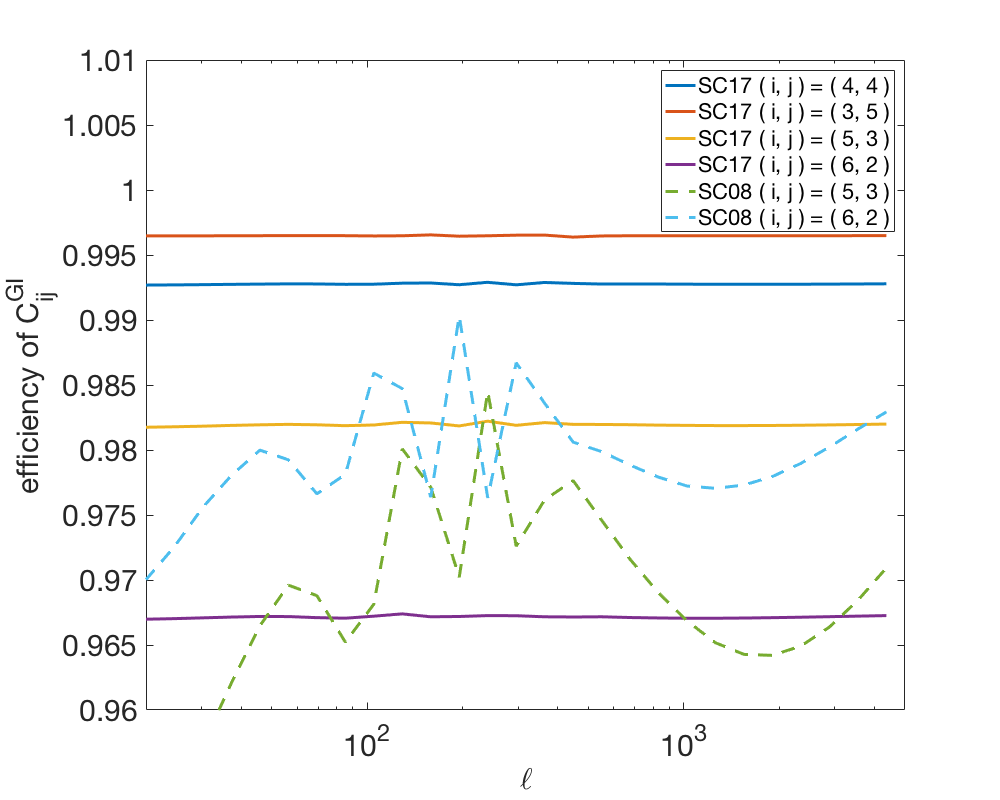}
		\caption{GI efficiency with different bin-pairs. Y-axis represents the efficiency, defined by ${C}^{GI(SC)}/{C}^{GI({\rm true})}$. X-axis represents the angular modes $\ell$. For all the bin-pairs and the total range of $\ell$, the accuracy is close to 1. $(i,j)=(4,4)$ has an efficiency very close to 1, reflecting the ability of this method to deal with GI in auto spectra, which is different from the SC2008 method \citep{Yao2017,SC2008}. $(i,j)=(3,5)$ shows that this method can also deal with ${C}^{GI}_{ij}$ with $i<j$ at good accuracy. $(i,j)=(5,3)$ shows this method can deal with ${C}^{GI}_{ij}$ with $i>j$ in the adjacent bin-pairs, but for larger separation the accuracy will drop quickly. See for example $(i,j)=(6,2)$.}
		\label{fig:GI_efficiency}
	\end{figure}

	Following the same procedure, we can obtain the other IA spectra, $\textit{C}^{GI}_{ij}$. Namely,  
	\begin{subequations}
		\begin{align} 
		\label{my-GI}
		\textit{C}^{GI}_{ij}(\ell)&\approx A^{Ig}_{ij}(\ell)\textit{C}^{Gg}_{ij}(\ell)~, \\
		\textit{C}^{Gg}_{ij}(\ell)&= \textit{C}^{(G+I)g}_{ij}(\ell)-\textit{C}^{Ig}_{ij}(\ell)~ \label{my-Gg}. 
		\end{align}
	\end{subequations}
    After obtaining the $A^{Ig}_{ij}$ as depicted in Eq.\,\eqref{steps1}, further steps can be applied as follows: 
	\begin{equation} \label{steps2}
	A^{Ig}_{ij} \stackrel{Eq.\,\eqref{my-Ig}}{\longrightarrow} C^{Ig}_{ij} \stackrel{Eq.\,\eqref{my-Gg}}{\longrightarrow} C^{Gg}_{ij} \stackrel{Eq.\,\eqref{my-GI}}{\longrightarrow} C^{GI}_{ij}.
	\end{equation}
	Here $\textit{C}^{(G+I)g}_{ij}(\ell)=\textit{C}^{(2)}_{ij}$ is the observed shape-galaxy spectra. $\textit{C}^{Ig}_{ij}$ is obtained from Eq.\,\eqref{my-Ig} and can be subtracted from $\textit{C}^{(G+I)g}_{ij}$. As shown in Fig.\,\ref{fig:GI_efficiency}, the efficiency of Eq.\,\eqref{my-GI} is accurate at the $1\%$ to $10\%$ level, depending on what bins are used. The efficiency is numerically calculated by dividing the SC2017-measured spectrum by the true GI spectrum. A value 1 represents the exact result from Eq.\,\eqref{steps2}.
	
	Basically, the efficiency is very good for $\textit{C}^{GI}_{ij}$ with $i\leq j$. The difference between the simulated spectra $\textit{C}^{GI}_{\rm (true)}$ and the SC obtained spectra $\textit{C}^{GI}_{\rm (SC)}$ is at the $1\%$ level. For $i>j$, the efficiency will drop quickly when we increase the separation between the redshift bins. However, for $i-j=3$ or $4$, the difference is still within $5\%$ level and the efficiency is still reasonably good. For a larger $i-j$ value, the efficiency will be significantly lower, and SC2017 does not remain comparable to SC2008 so we use SC2008 for such pairs of bins. 
    
For the SC2017 method, the extra observables being used include $C^{Ig}$ (which results from Eq.\,\eqref{scaling-2}, thus from $C^{(2)}$) and $C^{gg}$. But in the steps to propagate the information of IA, we did not use the statistical power of either galaxy-galaxy lensing or galaxy-galaxy clustering. They are only used for subtracting the IA signal and their measurement uncertainties will result in extra uncertainty in measuring the IA spectra, which we discuss in section \ref{s: error}.
	 
\subsection{IA and galaxy bias Model dependencies of SC methods}

It is worth clarifying to what extent SC methods are dependent or independent from IA and galaxy bias models. We delineate in the Appendix different cases of linear and non-linear IA or galaxy bias models for each SC method.

In summary, the self-calibration framework is not generically expandable to complete, higher-order IA models, as it was originally designed to work at linear order in the power spectrum. We denote the IA models that are linear in the matter power spectrum as linear IA models (but the non-linear matter power spectrum is used), while the higher-order IA models are denoted as non-linear IA models. It is possible to derive additional corrections at higher order for the approximation with a loss of efficiency in some cases. For the models typically considered, and all those at leading order, the method is IA model-independent. For higher-order term models, it is the structure of the galaxy bias that plays a determining role in how the SC methods are IA model dependent or not. The SC2008 is simply IA model independent in the case of linear bias (i.e. even for non-linear IA models). However, for non-linear bias, SC2008 acquires a term in its scaling relation, that is equal to $b_2 B_{I\delta\delta}/P_{I\delta}$, where $B_{I\delta\delta}$ is the bi-spectrum. Therefore, when using the SC2008 scaling relation in situations where non-linear bias and non-linear IA models are at play, one should expect a loss of efficiency in the method that needs to be estimated from theory and incorporated in the results due to neglecting this term.  On the other hand, the SC2017 is based on a relation that requires a linear bias model as well as a linear IA model and is thus model dependent.

	\section{Error Estimate for SC2017} \label{s: error}
	
	The SC2017 method for II (Section.\,\ref{s:SC II}) and GI (Section.\,\ref{s:SC GI}) has proven to be accurate because the differences between the SC estimations and the simulated values are small, so the residual (after cleaning IA using SC) will be small as well. However, introducing the extra observables $\textit{C}^{gg}_{ij}$ and $\textit{C}^{Ig}_{ii}$  creates extra measurement errors that need to be estimated and discussed.
	
	The measurement errors for the two introduced observables are given by:
	\begin{subequations}
		\begin{align} 
		\label{error Dgg}
		\Delta \textit{C}^{gg}_{ij}&=\sqrt{\frac{2}{(2\ell+1)\Delta\ell f_{\rm sky}}}\left(\textit{C}^{gg}_{ij}+\delta_{ij}\textit{C}^{gg,N}_{ii}\right)~,\\
		\label{error DIg}
		(\Delta \textit{C}^{Ig}_{ii})^2&=\frac{1}{2l\Delta l f_{\rm sky}} \left( {\textit{C}^{gg}_{ii}\textit{C}^{GG}_{ii}+\left[1+\frac{1}{3(1-Q)^2}\right]}\right.\notag\\
		&\times[\textit{C}^{gg}_{ii}\textit{C}^{GG,N}_{ii}+\textit{C}^{gg,N}_{ii}(\textit{C}^{GG}_{ii}+\textit{C}^{II}_{ii})] \notag\\
		&\left.{+\textit{C}^{gg,N}_{ii}\textit{C}^{GG,N}_{ii}\left[1+\frac{1}{(1-Q)^2}\right]}  \right)~.\\\notag
		\end{align}
	\end{subequations}
	Eq.\,\eqref{error Dgg} is the well known measurement error for gg spectra, including the effect of cosmic variance and shot noise. Eq.\,\eqref{error DIg} was derived  \citep{SC2008}. According to the steps introduced to self-calibrate the II and GI spectra, these two errors will also propagate to the II and GI spectra.
    \textcolor{black}{ We express the target power spectra in terms of the above extra observables through the steps of Eq.\,\eqref{steps1} and \eqref{steps2}:
\begin{subequations}
	\begin{align}
	C^{II}_{ij}&\approx\frac{C^{Ig}_{ii}C^{Ig}_{jj}}{C^{gg}_{ii}C^{gg}_{jj}}C^{gg}_{ij} ,\label{shortcut-1}\\
    C^{GI}_{ij}&\approx\sqrt{\frac{C^{Ig}_{ii}C^{Ig}_{jj}}{C^{gg}_{ii}C^{gg}_{jj}}}\left(C^{(2)}_{ij}-\sqrt{\frac{C^{Ig}_{ii}C^{Ig}_{jj}}{C^{gg}_{ii}C^{gg}_{jj}}}C^{gg}_{ij}\right). \label{shortcut-2}
	\end{align}
\end{subequations} }

	By propagating the error of Eq.\,\eqref{error Dgg} and Eq.\,\eqref{error DIg}, through the steps that have been introduced in the left hand side of Eq.\,\eqref{steps1} together with Eq.\,\eqref{steps2}, \textcolor{black}{or equivalently using Eq.\,\eqref{shortcut-1} and \eqref{shortcut-2}, while naively assuming no correlation between the included terms since the covariance will be extremely complicated (the impact of the covariance could only double the error as maximum, which will not affect the following results much),}
	%\begin{align} \label{steps3}
	%	A^{Ig}_{ij} \stackrel{Eq.\,\eqref{my-GI}}{\longrightarrow} \textit{C}^{GI}_{ij}~,
	%\end{align}
	the measurement error of this SC method can be obtained as $\Delta \textit{C}^{II}_{ij}$ and $\Delta \textit{C}^{GI}_{ij}$. The comparison can be made by calculating $\Delta \textit{C}^{II}_{ij}/\Delta \textit{C}^{GG}_{ij}$ and $\Delta \textit{C}^{GI}_{ij}/\Delta \textit{C}^{GG}_{ij}$ where $\Delta \textit{C}^{GG}_{ij}$ is the measurement error of cosmic shear, defined as:
	\begin{equation}
	\Delta \textit{C}^{GG}_{ij}(\ell)=\sqrt{\frac{2}{(2\ell+1)\Delta\ell f_{\rm sky}}}\left[\textit{C}^{GG}_{ij}(\ell)+\delta_{ij}\textit{C}^{GG,N}_{ii}\right]~.
	\end{equation}

	\begin{figure*}
		\includegraphics[width=2.0\columnwidth]{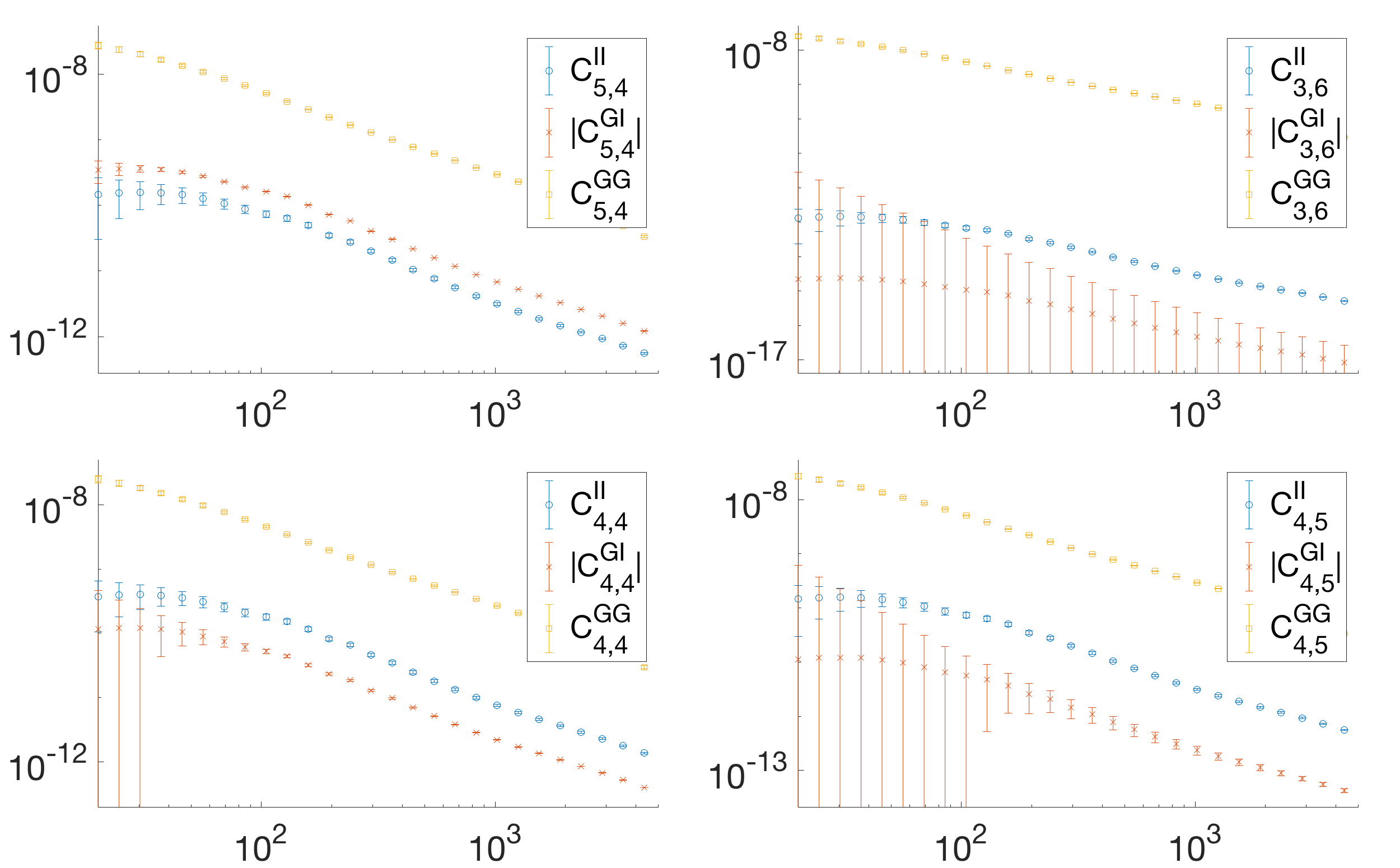}
		\caption{Numerical examples for the propagated errors for different spectra. The 4 panels use different redshift bin pairs $(i,j)$, which are $(4,4)$ on the bottom-left, $(4,5)$ on the bottom-right, $(5,4)$ on the top-left and $(3,6)$ on the top-right. Each panel includes the $C^{GG}$, $C^{GI}$ and $C^{II}$ signal. Some errorbars of $C^{GI}$ are large and the lower limit goes to negative. It means at those $(i,j,\ell)$ the S/N is low so that the SC2017 of GI could not apply. The main point of this figure is to show that the major contamination $C^{II}$ has good S/N, while the GI detection can only be applied to the auto-bins and adjacent-bins. The S/N for GI spectrum is good when its magnitude is not negligible compared to II spectrum.}
		\label{fig: errorplot}
	\end{figure*}
    
	\begin{figure}
		\includegraphics[width=1.1\columnwidth]{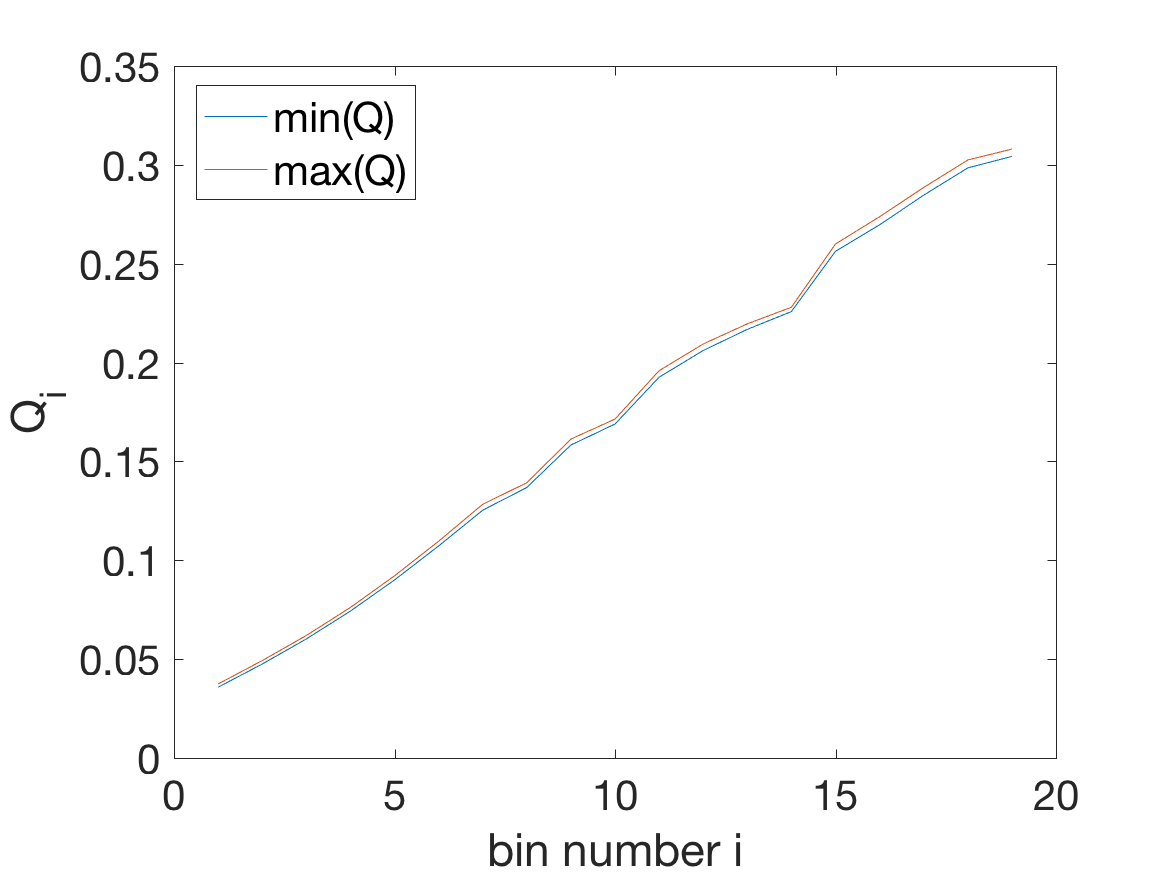}
		\caption{This figure presents the minimum and maximum values of $Q_i(\ell)$ in each redshift bin. In each bin, the minimum value and maximum value are very close, so that $Q_i(\ell)$ remains almost constant. The fact that values throughout all the redshift bins are not close to $1$ shows that the quality of photo-z is acceptable for SC to apply, without bringing much measurement uncertainty due to photo-z, see Eq.\,\eqref{error DIg}.}
		\label{fig: Qz}
	\end{figure}
    
	In this way, we numerically calculate the extra introduced error and compare with the error in cosmic shear, i.e. the fractional increase in the measurement error. The results are as follows:
	
	(1) For $\Delta \textit{C}^{II}_{ij}/\Delta \textit{C}^{GG}_{ij}$ throughout all the $\{i,j,\ell\}$ used, most values are at $<1\%$ level, with the overall average at $\sim 1\%$. Thus, the SC for II works well without introducing any significant extra error.
	
\begin{figure}
		\includegraphics[width=0.9\columnwidth]{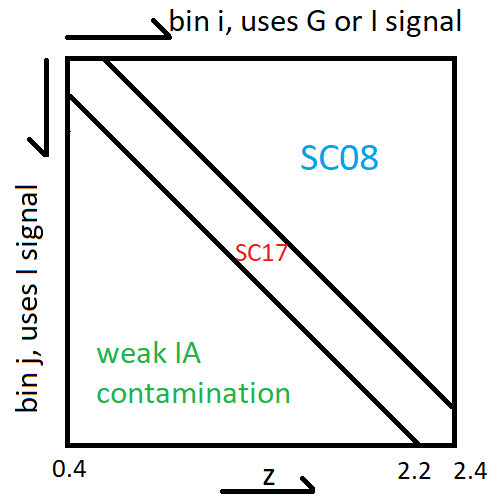}
		\caption{An illustration of how the combined SC works. The SC2008 only deals with $i<j$ for $C^{IG}_{ij}$, or $i>j$ for $C^{GI}_{ij}$ (the second notation is what we are using in this figure). With SC2008 cleaning the major contamination terms, we can have a much more precise estimation on the cosmological parameters, see \protect\cite{Yao2017}. The SC2017 method introduced in this work, can deal with most of the residual IA spectra after SC2008 is applied. It allows one to use the auto-spectra $i=j$. This figure shows the areas of the tomographic bin-pairs for which the SC methods are used. The blue SC08 area is where SC2008 is applied. The red SC17 area is where SC2017 is applied. The green area presents where both GI and II contamination are negligible. In the red SC17 area, SC2008 works poorly and has low efficiency. While in the blue SC08 area, SC2017 has large extra measurement errors so cannot be applied. This figure shows that SC2008 and SC2017 are two complementary methods. The choice of the boundary (black line) of these two methods depends on the efficiency of the methods depending on the redshift binning. With our binning method, the boundary line can be chosen to be $|i-j|\leq 1$ for SC2017, and $i-j>1$ for SC2008. We exaggerated the size of the red SC17 area in this figure for readability.}
		\label{fig SC17 illu}
	\end{figure}

	(2) For $\Delta \textit{C}^{GI}_{ij}/\Delta \textit{C}^{GG}_{ij}$ throughout all the $\{i,j,\ell\}$ used, most values are at $1\%\sim 10\%$ level, with an overall average of $\sim4\%$. Large values appear when the following three conditions arise: low $j$, high $i$ and low $\ell$. If we constrain $i\leq j+1$, the large fractional increase can be avoided, so that most values will be at $1\%\sim5\%$ level. In this case, a limited number of values reach $>10\%$. Fig.\,\ref{fig SC17 illu} illustrates the large error appearing in the SC08 area.

(3) We further test the detectability of the spectra and their errors in Fig.\,\ref{fig: errorplot}. The S/N of $C^{II}$ is very good in the major II contamination area (red 'SC17' in Fig.\,\ref{fig SC17 illu}). The GI contamination is shown to be generally lower than those of II. The detectability of GI is also good when the signal is significant, see top-left panel of Fig.\,\ref{fig: errorplot} with $(i,j)=(5,4)$.
    
(4) The estimated $Q_i(\ell)$ for each bin is shown in Fig.\,\ref{fig: Qz}. The $Q$ values remain almost constant in each redshift bin, which agrees with \cite{SC2008} and \cite{Yao2017}. The increasing pattern as we increase the redshift bin number $i$ is due to the photo-z scatter model $\sigma_z(1+z)$, see Eq.\,\eqref{redshift}. The value of $Q$ is smaller than discussed in \cite{Yao2017}, which is caused by the different $\sigma_z$ value applied. The fact that $Q<0.5$ is expected as discussed in \cite{SC2008} and assures that the photo-z error will not affect SC much through $Q$ and Eq.\,\eqref{scaling-2}. The measurement uncertainty of $Q_i$ is expected to be negligible as discussed in \citep{SC2008}.
    
	Through this analysis, it has been shown that SC2017 of II (Section.\,\ref{s:SC II}) and GI (Section.\,\ref{s:SC GI}) are not only accurate, but also introduce limited extra measurement errors in the auto- and adjacent- bins. Next, we will show that SC2017 offers a completion to circumvent the limitations of SC2008 discussed in Subsection.\,\ref{subsection SC2008}.
	
%	In SC2008, the contamination of $\textit{C}^{II}$ with $i\ne j$  and $\textit{C}^{GI}_{ij}$ with $i<j$ are minimized by the binning method, which yields the fact that SC2008 cannot deal with auto-spectra as the II signal is significant there. Also, SC2008 will have residual error in close bin-pairs. Finally, the small-bin approximation requires the choice of bin-width to be small enough. However, because of the presence of II signal, it cannot be too small, otherwise the effect of II in close bins will become significant. In previous works the choice of bin-width is $\Delta z^P=0.2$ assures the efficiency of the small-bin approximation while maintaining negligible II signal in bin-pairs.
	
	With the improvement of SC2017, $\textit{C}^{II}$ and $\textit{C}^{GI}_{ij}$ are taken care of in the auto- and adjacent- bin pairs where SC2008 works poorly, such that the auto-spectra can be used again to maximize lensing signal. Plus, the residual error in close bin-pairs is largely reduced. Finally, the limit of $\Delta z^P=0.2$ is no longer a problem, as the II signal is removed with a good accuracy. On the other hand, SC2008 can compensate for the limitation of SC2017 for the GI signal. SC2017 has a low efficiency and large extra measurement error for low $j$ and high $i$ values while SC2008 has very good efficiency with small extra measurement error for those bin-pairs. So these two methods can be used in a complementary way. 
	
	Finally, we want to point out that the improvement from using SC2017 is  based on the study of \cite{Zhang2010} and the small-bin approximation discussed in Eq.\,\eqref{my-II}, \eqref{my-Ig} and \eqref{my-GI}. These require strong correlation between the IA and the galaxy density fields meaning that that the 3-D power spectra should obey $(P^{Ig})^2=P^{II}P^{gg}$. From N-body simulations, the halo IA is indeed tightly correlated with the density field. But it has been shown that a mis-alignment exists between galaxies and the host halo. For LRGs, the mis-alignment angle is a random number with r.m.s of $\sim30\deg$ \citep{Okumura2009_LRG,Okumura2009,Chisari2017}. The presence of this mis-alignment angle can cause significant stochasticity between galaxy IA and galaxy density field, and can potentially weaken the approximations in Eqs.\,\eqref{steps1} and \eqref{steps2} by introducing an extra uncertainty to those equations. This should be studied in future work and is beyond the scope of this paper. 
	
	%%%%%%%%%%%%%%%%%%%%%%%%%%%%%%%%%%%%%%%%%%%%%%%%%%%%%%%%%%%%%%%%%%%
	%%%%%%%%%%%%%%%%%%%%%%%%%%%%%%%%%%%%%%%%%%%%%%%%%%%%%%%%%%%%%%%%%%%
	\section{Models and tools for application to cosmic shear surveys}
	\label{section models}
	%%%%%%%%%%%%%%%%%%%%%%%%%%%%%%%%%%%%%%%%%%%%%%%%%%%%%%%%%%%%%%%%%%%
	%%%%%%%%%%%%%%%%%%%%%%%%%%%%%%%%%%%%%%%%%%%%%%%%%%%%%%%%%%%%%%%%%%%
	
	In this work, we chose to use the survey specifications of LSST to show the improvement realized by SC2017 when compared to SC2008. The applications of SC2008 to LSST, Euclid and WFIRST were discussed in \cite{Yao2017}. In this section, we will describe models of photo-z, galaxy bias and IA being used. The IA models are used to calculate spectra to do forecasts, but the SC method does not use any of the IA models. The information for IA ($\textit{C}^{IG}$) completely comes from $\textit{C}^{Ig}$, which is obtained by Eq.~(\ref{scaling-2}), utilizing the information from photo-z. The fiducial cosmological model is described in Tab.\,\ref{table FiducialModel}.
	
	\begin{table}
		\centering
		\caption{Fiducial Cosmological Model}\label{table FiducialModel}
		%	\begin{ruledtabular}
		\begin{tabular}{ c c c c c c c }
			\hline
			$\Omega_m$ & $h_0$ & $\sigma_8$ & $n_s$ & $\Omega_b$ & $w_0$ & $w_a$ \\
			\hline
			0.315  &  0.673  &  0.829  &  0.9603  & 0.049 &  -1.0  &  0 \\
			\hline
		\end{tabular}
		%	\end{ruledtabular}
	\end{table}
	
	\subsection{Photo-z Model}
	
	Photo-z poses another major problem in future surveys \citep{Ma2006photoz}. The $\textit{C}^{Ig}_{ii}$ measurement also depends on the quality of photo-zs. Therefore, it is important to take such a difficulty into consideration. The overall true-z distribution of the survey and photo-z (Gaussian) probability distribution function are expressed as:
	\begin{align} 
	n(z)&\propto z^\alpha {\rm exp}\left[-(\frac{z}{z_0})^\beta\right],\\
	p(z^P|z)&=\frac{1}{\sqrt{2\pi}\sigma_z(1+z)}{\rm exp}\left[-\frac{(z-z^P-\Delta_z^i)^2}{2(\sigma_z(1+z))^2}\right].\label{redshift}
	\end{align}
	
	Here $z$ is the true-z, $z^P$ is the photo-z, $n(z)$ gives the overall redshift distribution of the survey, and $p(z^P|z)$ is the probability distribution function (PDF) of photo-z for a given true redshift. The normalized redshift distribution for each tomographic bin is expressed as $n_i(z)$, which is given by:
	\begin{align} \label{n_i}
	n_i(z)=\frac{\int_{z_{i,\rm min}^P}^{z_{i,\rm max}^P} n(z)p(z^P|z)dz^P}{\int_0^\infty\left[\int_{z_{i,\rm min}^P}^{z_{i,\rm max}^P} n(z)p(z^P|z)dz^P\right]dz}.
	\end{align}
	
	As mentioned earlier, in this work we chose the specifications for  LSST \citep{DESC2012,Chang2013}. In order to show the full extent of improvements possible from SC2017, we use smaller photo-z bin-width and chose an optimistic LSST photo-z scatter of $\sigma_z=0.025$ \citep{Tyson2017}, which is half of the fiducial value $\sigma_z=0.05$ that we used in our previous forecast work using only SC2008~\citep{Yao2017}. This can be made possible with the development in photo-z techniques such as machine learning photo-z \citep{Bilicki2017}, as well as including synergy with other surveys such as H band from WFIRST \citep{Tyson2017}. The redshift bias is chosen as $\Delta_z=0$, with photo-z priors chosen to be $\sigma_z={\rm Gaussian}(0.025,0.005)$ and $\Delta_z={\rm Gaussian}(0,0.006)$. The survey specifications are shown in Table.\,\ref{tab surveys}.
	
	\begin{table}
		\centering
		\caption{Survey Parameters \citep{DESC2012,Chang2013}}\label{tab surveys}
		%	\begin{ruledtabular}
		\begin{tabular}{ c c c c c c c c }
			\hline
			& $f_{\rm sky}$ & $\gamma_{\rm rms}$ & $n_{\rm eff}$ & $z_0$ & $\alpha$ & $\beta$ & $z_{\rm max}$ \\
			\hline
			LSST & 0.436 & 0.26 & 26 & 0.5 & 1.27 & 1.02 & 3.5 \\
			\hline
		\end{tabular}
		%	\end{ruledtabular}
	\end{table}
	
\begin{figure}
		\includegraphics[width=0.9\columnwidth]{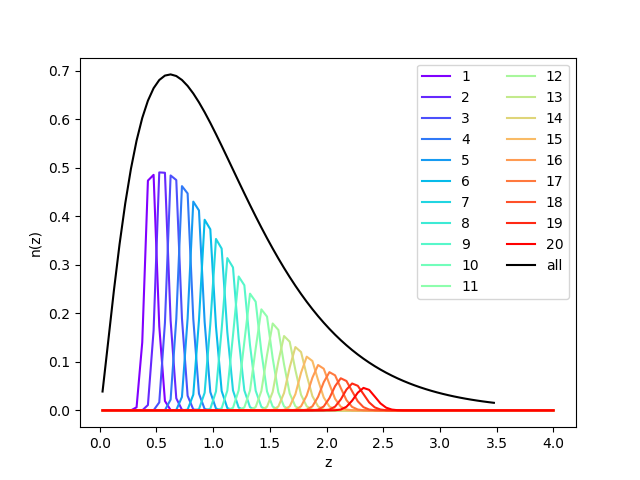}
		\caption{Redshift distribution with 20 tomographic bins of SC. Photo-z range $[0.4,2.4)$ with bin-width $\Delta z^P=0.1$. In this plot the redshift distributions are not normalized, so they are calculated by the numerator of Eq.~\eqref{n_i}. This is for convenient comparison with the overall distribution.
			\label{fig n(z)}}
	\end{figure}

	In our previous work \cite{Yao2017}, the choice of the redshift binning method for SC2008 was carefully discussed. We used a redshift range of $[0.4,2.4)$, a bin-width $\Delta z^P=0.2$ and an overall bin-number of $10$. In this work, the purpose of SC2017 is to push the SC method to a smaller redshift bin-width beyond the limit of SC2008 while improving the efficiency and reducing the residual bias. Thus, we chose the overall bin-number to be $n_{\rm bin}=20$ with a bin-width of $\Delta z^P=0.1$.
	Fig.~\ref{fig n(z)} shows the unnormalized redshift distribution of 20 tomographic bins for LSST with photo-z range $0.4\leq z^P<2.4$. %In our previous forecast work \citep{Yao2017}, we also explored the effort of SC2008 for WFIRST and Euclid with their survey specification.
	
	\subsection{Bias Model}
	
	The actual SC technique uses a bias measured from using ${C}^{gg}_{ii}(l)\approx b_i^2(l){C}^{mm}_{ii}(l)$ for real data, where ${C}^{mm}$ comes from the CMB experiments as discussed in \cite{SC2008}. Also, as discussed in \cite{Yao2017}, the CMB will give a set of tightly constrained best-fit cosmological parameters. These cosmological parameters constraints will not deviate significantly from the expected values of cosmic shear, so that they can be chosen as fiducial cosmology for the SC method at this point, although without using the statistical power of other experiments. Furthermore, for a more precise late-time cosmology one can apply CMB+SN+BAO to get even better constrained cosmological parameters. In this work, we assess the bias by averaging over each redshift bin as before \cite{Yao2017}. That is $b_i=\int_0^\infty b_g n_i dz$ which requires a bias model, $b_g$.
	
	We use the Generalized Time Dependent (GTD) Bias Model \citep{Clerkin2014}, which is an encapsulation of several time-dependent models and agrees with simulations \citep{Clerkin2014}. The expression for the GTD model is given by:
	\begin{align}
	b(z)=c+(b_0-c)/D^\alpha(z) \label{GTD model},
	\end{align}
	in which $D(z)$ is the linear growth function, satisfying 
    \begin{equation}
    \ddot{D}+2H(z)\dot{D}-\frac{3}{2}\Omega_m H_0^2(1+z)^3D=0.
    \end{equation}
    The parameters we use are $c=0.57$, $b_0=0.79$ and $\alpha=2.23$ \citep{Clerkin2014}.
	
	The scaling relation Eq.~(\ref{scaling-1}) of SC is not sensitive to the choice of bias model. This is because the galaxy bias enters both $\textit{C}^{Ig}$ and $b_i$, hence they roughly cancel out each other.
    
We want to point out that for the bias model used, the efficiency of SC2017 is practically scale-independent, see Fig.\,\ref{fig II efficiency} and \ref{fig:GI_efficiency}. For a scale-dependent galaxy bias model, the efficiency would also be affected and is expected to be scale-dependent. This is further discussed in the Appendix \ref{appendix}.
	
	\subsection{IA model (used to generate forecast spectra)}
	
	The components in Eq.~(\ref{observables}) are given by, e.g.  \cite{BridleKing} 
	\begin{subequations}
		\begin{align}
		\textit{C}^{GG}_{ij}(\ell)&=\int_0^\infty\frac{q_i(\chi)q_j(\chi)}{\chi^2}P_\delta\left(k=\frac{\ell}{\chi};\chi\right)d\chi, \\ \label{GG}
		\textit{C}^{IG}_{ij}(\ell)&=\int_0^\infty\frac{n_i(\chi)q_j(\chi)}{\chi^2}P_{\delta,\gamma^I}\left(k=\frac{\ell}{\chi};\chi\right)d\chi, \\ \label{IG}
		\textit{C}^{GI}_{ij}(\ell)&=\int_0^\infty\frac{q_i(\chi)n_j(\chi)}{\chi^2}P_{\delta,\gamma^I}\left(k=\frac{\ell}{\chi};\chi\right)d\chi, \\ \label{GI}
		\textit{C}^{II}_{ij}(\ell)&=\int_0^\infty\frac{n_i(\chi)n_j(\chi)}{\chi^2}P_{\gamma^I}\left(k=\frac{\ell}{\chi};\chi\right)d\chi. \\
		\end{align}
	\end{subequations}
	
	$P_\delta(k;\chi)$ in Eq.~(\ref{GG}) is simply the (non-linear) matter power spectrum at the redshift of the lens. $n_i(\chi)$ is the redshift distribution in the i-th redshift bin. $q_i(\chi)$ is the lensing efficiency function for lens at $\chi_L(z_L)$ for the i-th redshift bin, written as
	\begin{equation}
	q_i(\chi_L)=\frac{3}{2}\Omega_m\frac{H_0^2}{c^2}(1+z_L) \int_{\chi_L}^\infty n_i(\chi_S)\frac{(\chi_S-\chi_L)\chi_L}{\chi_S}d\chi_S.
	\end{equation}
	
	The two relevant 3-D IA spectra are
	\begin{subequations}
		\begin{align} \label{IA 3D}
		P_{\delta,\gamma^I}&=-A(L,z)\frac{C_1\rho_{m,0}}{D(z)}P_\delta(k;\chi), \\
		P_{\gamma^I}&=A^2(L,z)\left(\frac{C_1\rho_{m,0}}{D(z)}\right)^2P_\delta(k;\chi). 
		\end{align}
	\end{subequations}
	
	In Eq.~(\ref{IA 3D}), $\rho_{m,0}=\rho_{crit}\Omega_{m,0}$ is the mean matter density of the universe at $z=0$. $C_1=5\times 10^{-14}(h^2M_{\rm sun}/Mpc^{-3})$ was used in \cite{BridleKing}. We use $C_1\rho_{crit}\approx 0.0134$ as in \cite{Krause2016}.
	
	Here we use the redshift- and luminosity dependent IA amplitude model $A(L,z)$ obeying the following redshift and luminosity scaling
	\begin{equation}
	A(L,z)=A_0\left(\frac{L}{L_0}\right)^\beta\left(\frac{1+z}{1+z_0}\right)^\eta.
	\end{equation}
	
	The IA parameters for this model are $A_0$, $\beta$ and $\eta$, with $z_0=0.3$ the (observationally motivated) pivot redshift \citep{Krause2016}, and $L_0$ (the pivot luminosity) corresponding to an absolute magnitude of -22 in r-band.
	
	In this work, the IA amplitude is chosen to be $A_0=1$. We fix $\beta=0$ so the model is luminosity-independent in the presented results. This assumption is reasonable for the purpose of this work as it has been shown that in CFHTLenS data \citep{Joudaki2016} and KiDS-450 data \citep{Hildebrandt2016}, the IA signal is insensitive to any luminosity-dependence with the full lensing catalog of their surveys, although it may not be the case for an LSST-like survey \citep{Krause2016}. Indeed, it has been shown in \cite{Joachimi2011,Singh2015} that the IA is dependent on the luminosity. However, we assume here that the SC is mainly not sensitive to the specific underlying IA signal, as the major contamination is cleaned by the SC2008 component, which is mainly IA-model independent, see Appendix \ref{appendix}. The redshift-dependency $\eta$ is chosen to be 0 for the same reason. We therefore use the model introduced above. In future works we will explore the effect of a more complicated IA model with a varying $\beta$, as well as the impacts of different luminosity functions of red- and blue-type galaxies. As discussed before, these differences in the IA model could potentially weaken the efficiency of SC2017, however the impact is not expected to be significant, as the dependence on the IA model is weak and the differences can be smoothed out due to the Limber integrals (Eqns.\,\eqref{II}, \eqref{Ig} and \eqref{gg}).
    
We did not include the impact of one-halo IA regime \citep{Schneider2010,TroxelIshak} even if the redshift range and angular $\ell$ range we use do reach scales with $k\sim 1$ h/Mpc. The measured Ig spectrum already contains this information and the principal component of the self-calibration, i.e. SC2008, is mainly model-independent so we expect that this shortcoming will only weaken the efficiency of the SC2010-based part of the overall method. The effect of one-halo model is thus expected here to only have a weak effect in a limited redshift bins and a limited $\ell$ range. 
	    
	\begin{figure*}
		\includegraphics[width=2.3\columnwidth]{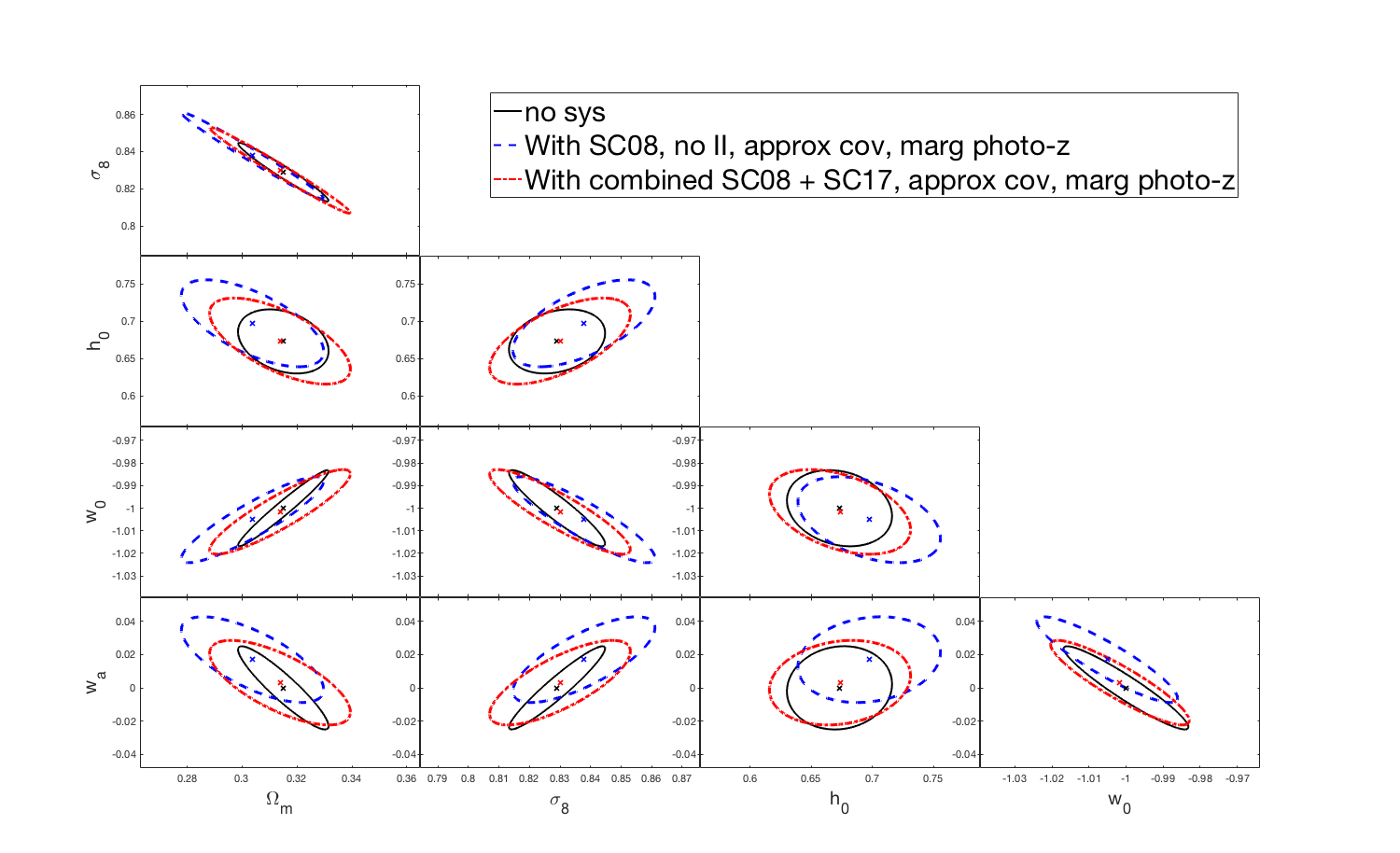}
		\caption{Confidence contours for the forecast using the combined SC method. The (black) ``no sys'' one stands for no systematics scenario, where there is no IA in the simulated spectra, and photo-z is assumed to be perfect so there is no need for marginalization. (This is the perfect reference case.) The blue one uses the SC 2008 to clean the IA contamination in the forecast, with the presence of both GI and II in the simulated spectra, marginalized over photo-z parameters. The red uses the combined SC method, with presence of both GI and II in the simulated spectra, marginalized over photo-z parameters. In our previous work, we showed the improvement from full IA contamination to the shifted position after SC2008 is applied, while this figure shows the improvement from the actual performance of SC2008 to the case of applying SC2008 + SC2017. The major improvement is that SC2017 cleans most of the residual error from SC2008. For comparison between SC2008 and the marginalization method, we refer the reader to our previous work \protect\cite{Yao2017}.}
        \label{fig shift_new_SC}
	\end{figure*}
	
	\subsection{Shift of Best-fit Cosmological Parameters}
	
	In our previous work, we presented the shift of the best-fit cosmological parameters biased by the residual signal after SC is applied. This shift has been shown to be very small, and the corresponding best-fit values are very close to the fiducial values, which is a good improvement compared to the case with full IA contamination. In this work, we will again apply the Newtonian method that has been used to calculate the residual shift to show that the new SC2017 method will provide further imporvements to the original SC2008 method.
	
	The shift is calculated by
	\begin{equation} \label{Newtonian shift}
	\Delta P^\beta\approx-(F^{-1})^{\alpha\beta}\mathcal{A}_\alpha,
	\end{equation}
	where $\Delta P^\beta=P^\beta|_{\rm CM2}-P^\beta|_{\rm CM1}$ is the shift from fiducial cosmology CM1 (cosmological model 1) to the best-fit cosmology CM2 (cosmological model 2), $F$ is the Fisher matrix and $\mathcal{A}_\alpha$ is defined as the partial derivative of the likelihood:
	\begin{align}
	\mathcal{A}_\alpha&=(\frac{1}{2}\chi^2)_{,\alpha}=\frac{1}{2}\frac{\partial \chi^2}{\partial p^\alpha}\\
	&=f_{\rm sky}\sum\limits_{l}\frac{(2l+1)\Delta l}{2}\left[\textit{C}^{GG}-\textit{C}^{GG(SC)}\right]_{ij} \notag \\
	&\times\left[Cov(\textit{C}^{GG(SC)},\textit{C}^{GG(SC)})\right]^{-1}_{(ij),(pq)}\frac{\partial \textit{C}^{GG}_{(pq)}}{\partial p^\alpha} \label{A_alpha}~.\notag
	\end{align}
	
	From the expression of $\mathcal{A}_\alpha$ we can see that the source of the shift comes from this term $\left[\textit{C}^{GG}-\textit{C}^{GG(SC)}\right]_{ij}$, which is the difference between the true signal and the SC-measured signal. As the efficiency of SC is improved by the SC2017 method, we are expecting to see a significantly smaller shift, which will be shown in the next section. The detailed derivation of this Newtonian method to calculate the shift was discussed in our previous work \cite{Yao2017} and others \cite{Kirk2012,Clerkin2014}.
	
	%%%%%%%%%%%%%%%%%%%%%%%%%%%%%%%%%%%%%%%%%%%%%%%%%%%%%%%%%%%%%%%%%%%%%%%%%%%%%%%%%%%%%%%%%
	%%%%%%%%%%%%%%%%%%%%%%%%%%%%%%%%%%%%%%%%%%%%%%%%%%%%%%%%%%%%%%%%%%%%%%%%%%%%%%%%%%%%%%%%
	\section{Results and cosmological parameter forecasts from applying the Combined SC Methods} \label{section results}
	%%%%%%%%%%%%%%%%%%%%%%%%%%%%%%%%%%%%%%%%%%%%%%%%%%%%%%%%%%%%%%%%%%%%%%%%%%%%%%%
	%%%%%%%%%%%%%%%%%%%%%%%%%%%%%%%%%%%%%%%%%%%%%%%%%%%%%%%%%%%%%%%%%%%%%%%%%%%%%%%
	
	We present a comparison between the SC2008 method and the combined SC2008 + SC2017 method.  The SC2017 is applied in the red SC17 area in Fig.\,\ref{fig SC17 illu}. SC2008 is applied to the major GI contaminated area $i-j>1$, while SC2017 is applied to the major II contaminated area $|i-j|\leq 1$.
	
	\begin{table}
		\centering
		\caption{We found in our previous paper \protect\cite{Yao2017} that SC2008 reduces the shift in cosmological parameters from several-$\sigma$ to below 1-$\sigma$.  We find here that SC2017 makes further improvements in reducing the shifts.  We use $\Delta p^{\rm SC17}$ for the residual shift of SC2017 for each parameter, while $\Delta p^{\rm SC08}$ is for the residual shift of SC2008 for each parameter. The ratios show how the SC methods reduce the bias below the $1-\sigma$ and that SC2017 makes even further improvements on such reduction. } 
		\label{tab reduced shift}
		%	\begin{ruledtabular}
		\begin{tabular}{ c  c  c  c  c  c  }
			\hline
			values & $\Omega_m$ & $h_0$ & $\sigma_8$ & $w_0$ & $w_a$ \\
			\hline
			%$\Delta p^{\rm SC17}/\Delta p^{\rm SC08}$ & $0.13$ & $0.02$ & $0.13$ & $0.60$ & $0.27$ \\
            $\Delta p^{\rm SC08}/\sigma^{\rm SC08}$ & 0.56 & 0.61 & 0.47 & 0.28 & 0.83 \\
            $\Delta p^{\rm SC17}/\sigma^{\rm SC17}$ & 0.07 & 0.01 & 0.06 & 0.17 & 0.23 \\
			\hline	
		\end{tabular}
		%	\end{ruledtabular}
	\end{table}
	
	 Fig.\,\ref{fig shift_new_SC} shows a comparison between the SC2008 performance and SC2008+SC2017methods. Because of the presence of the II signals, the SC2008 blue contours shift to the opposite direction from our previous results in \cite{Yao2017}. This shift was not significant in our previous work, because the II signal as a major source of the shift was not taken into account by the choice of redshift bin-width of $\Delta z^P=0.2$. In the future, when a smaller bin-width is required, such as $\Delta z^P=0.1$, the shift will be significantly larger if SC2008 alone is applied. By applying the SC2017 method, most of the  ${C}^{II}$ and ${C}^{GI}$ contaminations are mitigated, bringing the best-fit cosmological parameters significantly closer to the fiducial values. The $1\sigma$ uncertainties are also reduced by about $1\%\sim2\%$ level due to the fact that the SC2017 now allows us to use the auto-spectra signal. This improvement is not significant because the number of auto-spectra is much less than the cross-spectra, so the impact of the extra constraining power by using the auto-spectra is small but the real gain is in reducing the residual shift is the parameters. The improvement in the best-fit cosmological parameters are numerically shown in Table.\,\ref{tab reduced shift}.     
    Although $\Delta p^{\rm SC17}/\Delta p^{\rm SC08}$ reflects improvements for all parameters, it appears that the improvements for $h_0$ is much larger than for other parameters. We verified that the results of the Fisher formalism converges properly by changing the step size in the numerical partial derivatives. Although, it is expected that changes in the cosmological parameter shifts are uneven due to the non-linear multi-dimensional space of cosmological parameters, this point should be explored further using MCMC analyses.
    
It is maybe useful to say a few words about how SC compares with the marginalization method. In our previous paper \cite{Yao2017}, we found that SC2008 is competitive compared with the marginalization method as both provide almost similar improvement in removing the bias in cosmological parameters. 
Also, the uncertainties caused by the two methods were also found comparable. The further improvement shown in Fig.\,\ref{fig shift_new_SC} suggests that SC2017 method can provide even further improvements although the new portion of the SC is not model independent.
	
	\section{Summary} \label{section summary}
	
	In this work we proposed a new SC method (SC2017) by combining strategies from other previous SC methods. In doing so, we applied the IA information in $\textit{C}^{Ig}_{ii}$ (measured by SC2008) into $\textit{C}^{II}_{ij}$ and $\textit{C}^{GI}_{ij}$ measurement by using modified relations based on the SC2010 method. In this way, we maximized the usage of the IA information in $\textit{C}^{Ig}_{ii}$ without strong assumption on the features of the IA model. 
    
	The SC2017 method was shown to add significant improvement to the residual shift in cosmological parameters compared to the SC2008 method. The reason is that for the auto-spectra or adjacent cross-spectra, SC2008 cannot mitigate the $\textit{C}^{II}$ and has low efficiency in $\textit{C}^{IG}$ measurements. By addressing these limitations, SC2017 not only allows one to obtain more accurate best-fit values for the cosmological parameters, but also makes the constraining power in the auto-spectra available. Furthermore, the SC2017 method allows one to use smaller redshift bin-width which allows SC methods to work even more efficiently.
	
	A point worth exploring in the future is what kind of uncertainties can be introduced in SC methods when taking into account the mis-alignment angle between galaxy and the host halo found in of N-body simulations. This is out of the scope of this paper and is left for future work. Also, the effect of non-linear IA and non-linear galaxy bias (see \cite{Blazek2015,Blazek2017} for example) can potentially lower the efficiency of the SC method, in spite of the small-bin-approximation being applied. Thus modifications similar to for example that of \cite{Troxel2012} will be needed for current SC methods. This is also beyond the scope of this paper but future work should be dedicated to this.
    
    We also want to point out that it is possible to mitigate IA in the close bin pairs using another method, since SC2017 does require some assumptions about the underlying IA model as discussed. One possible method is to measure the $C^{GI}$ spectra using the SC2008 method, then use it for IA model selection. The marginalization method can then be applied to clean the close bin pairs.
	
	With the improvements found in the SC methods, it becomes an appealing endeavor to apply them to mock catalogs or current and future photo-z surveys (e.g. CFHTLenS, KIDS, DES, LSST, and WFIRST) and that will be the subject of a follow up work. 
	
	%\begin{acknowledgements}
	\section*{Acknowledgements}
	
   This paper has undergone internal review in the LSST Dark Energy Science Collaboration. % REQUIRED
The internal reviewers were Elisa Chisari, Danielle Leonard, and Sukhdeep Singh. %\ldots. % Optional but recommended
J.Y. conducted the main analysis and writing of the manuscript; M.I. supervised and contributed to the analysis and the writing of the manuscript; M.T. participated in discussions during the analysis and paper writing.
 We thank W. Lin and P. Zhang for useful comments. We also thank L. Fox and L. Rayborn for proofreading the manuscript. M.I. acknowledges that this material is based upon work supported in part by NSF under grant AST-1517768, the U.S. Department of Energy, Office of Science, under Award Number DE-SC0019206, and an award from the John Templeton Foundation.
% Standard papers only: author contribution statements. For examples, see http://blogs.nature.com/nautilus/2007/11/post_12.html
% This work used TBD kindly provided by Not-A-DESC Member and benefitted from comments by Another Non-DESC person.
% Standard papers only: A.B.C. acknowledges support from grant 1234 from ...
% \input{desc-tex/ack/standard}
% This work used some telescope which is operated/funded by some agency or consortium or foundation ...
We acknowledge the use of CosmoSIS for parts of the analysis (\url{https://bitbucket.org/joezuntz/cosmosis/wiki/Home}).

% LSST-DESC acknowledgement https://github.com/LSSTDESC/desc-tex/tree/master/ack
The DESC acknowledges ongoing support from the Institut National de Physique Nucl\'eaire et de Physique des Particules in France; the Science \& Technology Facilities Council in the United Kingdom; and the Department of Energy, the National Science Foundation, and the LSST Corporation in the United States.  DESC uses resources of the IN2P3 Computing Center (CC-IN2P3--Lyon/Villeurbanne - France) funded by the Centre National de la Recherche Scientifique; the National Energy Research Scientific Computing Center, a DOE Office of Science User Facility supported by the Office of Science of the U.S.\ Department of Energy under Contract No.\ DE-AC02-05CH11231; STFC DiRAC HPC Facilities, funded by UK BIS National E-infrastructure capital grants; and the UK particle physics grid, supported by the GridPP Collaboration.  This work was performed in part under DOE Contract DE-AC02-76SF00515.	
	%\end{acknowledgements}

	\bibliographystyle{mnras}
	%\bibliography{example} % if your bibtex file is called example.bib
	\bibliography{references}

\appendix
\section{The Model Dependency of the SC Methods} \label{appendix}
	Aside from what we found for SC2017 as an improvement to SC2008, we would like to add this discussion to summarize the IA-model dependencies of SC methods of \cite{SC2008,Zhang2010,TroxelIshak,Yao2017} in view of recent IA models \citep{Blazek2015,Blazek2017}. We will separate the discussions in the conditions of linear/non-linear IA, but also for linear/non-linear galaxy bias. We do not include the discussion when both IA and galaxy bias are linear as both SC2008 and SC2017 are invariant when moving between different linear models. The discussion will be focused on the impact of working with non-linear models.
	
	\subsection{SC2008, linear IA and non-linear  galaxy bias case}
	
	We discussed in our previous work \cite{Yao2017} that the efficiency of SC2008 method will depend on linear galaxy bias, while for non-linear galaxy bias, the scaling relation of Eq.\,\eqref{scaling-1} will need to be adjusted. Here we present a more detailed discussion.
	
	We re-derive how SC2008 is obtained while assuming a bias model of the form:
	\begin{equation}
	\delta_g=b_{g,1}\delta_m+b_{g,2}\delta_m^2+O(\delta_m^3), \label{NL_bias}.
	\end{equation}
     The IA spectra read:
	\begin{align}
	C^{IG}_{ij}(\ell)&=\int_{0}^{\infty}\frac{n_i(\chi)W_j(\chi)}{\chi^2} P_{I\delta}\left(k=\frac{\ell}{\chi};\chi\right) d\chi , \\
	C^{Ig}_{ii}(\ell)&=\int_{0}^{\infty}\frac{n_i(\chi)n_i(\chi)}{\chi^2} P_{Ig}\left(k=\frac{\ell}{\chi};\chi\right) d\chi . \label{Ig_append}
	\end{align}
	Here in Eq.\,\eqref{Ig_append}, the intrinsic - galaxy correlation and the associated 3D power spectrum are expressed as:
	\begin{align}
	<\gamma^I\delta_g>&=<\gamma^Ib_{g,1}\delta_m>+<\gamma^Ib_{g,2}\delta_m^2>, \\
    P_{Ig}&=b_{g,1}P_{I\delta}+b_{g,2}B_{I\delta\delta}, \label{expand Ig}
	\end{align}
    in which $\gamma^I$ is the intrinsic ellipticity of a galaxy, $B_{I\delta\delta}$ is the intrinsic - matter - matter bi-spectrum.
	
	Due to the small bin approximation that we use, the above integrals can be approximated to:
	\begin{align}
	C^{IG}_{ij}(\ell)&\approx \frac{W_{ij}}{\chi_i^2}P_{I\delta}(k_i=\frac{\ell}{\chi_i};\chi_i) \label{IG limber},\\
	C^{Ig}_{ii}(\ell)&\approx \frac{1}{\chi_i^2\Delta_i}\left[b_{i,1}(\ell)P_{I\delta}+b_{i,2}B_{I\delta\delta}\right] . \label{Ig limber}
	\end{align}
	For the scaling relation Eq.\,\eqref{scaling-1} shown in previous studies \cite{SC2008,Yao2017}, the impact of non-linear galaxy bias (2nd term on the RHS of Eq.\,\eqref{Ig limber}) is ignored. For the correction including this effect, the corrected scaling relation will be:
	\begin{equation}
	C^{IG}_{ij}(l)\simeq \frac{W_{ij}\Delta_i}{b_{i,1}(l)+b_{i,2}\frac{B_{I\delta\delta}}{P_{I\delta}}}C^{Ig}_{ii}(l). \label{scaling-3}
	\end{equation}
	In this case, since we are considering linear IA model, $B_{I\delta\delta}\propto B_{\delta\delta\delta}$, which is the matter bi-spectrum, is zero if non-Gaussianity in the density field is ignored. Thus in this case SC2008 is model-independent.
	
	\subsection{SC2008, non-linear IA and linear galaxy bias case}
	
	In this case, we ignore the non-linear galaxy bias in Eq.\,\eqref{NL_bias}, but consider IA models beyond the linear case, for example the models for $P_{I\delta}$ from \cite{Blazek2015,Blazek2017}. It is obvious that the 3D IA spectra $P_{I\delta}$ will still get canceled in Eq.\,\eqref{IG limber} and \eqref{Ig limber}, but with $b_{i,2}=0$ in Eq.\,\eqref{Ig limber} and \eqref{scaling-3}. Thus this case is still IA-model independent.
	
	\subsection{SC2008, non-linear IA and non-linear galaxy bias case}
	
	Based on the previous discussions, the scaling relation will then be corrected to Eq.\,\eqref{scaling-3}. Due to the existence of the non-linear bias and the ratio $B_{I\delta\delta}/P_{I\delta}$, in which the IA model is not directly canceled, the efficiency of Eq.\,\eqref{scaling-3} will then be IA-model-dependent. In future work, it might be possible to derive additional corrections at higher order for the approximation.

	\subsection{SC2017, linear IA and linear galaxy bias case}
	
	The SC2017 method is based on SC2010, while making different assumptions from SC2008. For SC2017 to work, according to:
	\begin{align} 
	C^{II}_{ij}(\ell)&=\int_{0}^{\infty}\frac{n_i(\chi)n_j(\chi)}{\chi^2} P_{II}\left(k=\frac{\ell}{\chi};\chi\right) d\chi, \\
	C^{Ig}_{ij}(\ell)&=\int_{0}^{\infty}\frac{n_i(\chi)n_j(\chi)}{\chi^2} P_{Ig}\left(k=\frac{\ell}{\chi};\chi\right) d\chi, \\
	C^{gg}_{ij}(\ell)&=\int_{0}^{\infty}\frac{n_i(\chi)n_j(\chi)}{\chi^2} P_{gg}\left(k=\frac{\ell}{\chi};\chi\right) d\chi,
	\end{align}
	the requirement is that 
	\begin{equation}
	P_{II}P_{gg}\approx P_{Ig}^2. \label{scaling-4}
	\end{equation}
	For this relation to equal exactly, the linear IA and linear galaxy bias models are required, and hence SC2017 is model-dependent. For non-linear models, there will be an efficiency drop in Eq.\,\eqref{scaling-4}. For example, considering a complete IA model with higher order terms \citep{Blazek2017}, when expanding the expression of Eq.\,\eqref{scaling-4}, some terms on the RHS and LHS are exactly equal, however some terms have different dependencies, such as $(A_{0|0E}+C_{0|0E})^2$ on one side while $A_{0E|0E}P_\delta$ on the other. Here terms like $A_{0|0E}$ are correlation functions introduced in \cite{Blazek2017}.
    
    Therefore, from a theoretical point of view, SC2017 is a model-dependent method. The detailed study of how good the relation of Eq.\,\eqref{scaling-4} holds considering both higher order IA and higher order galaxy bias remains to be explored in future study. Recently a study \citep{Meng2018} in N-body simulation using SC2010 \citep{Zhang2010} showed that this method is not sensitive to the underlying IA signal. It is encouraging for the future study of SC2017, which is based on SC2010.
	
	\subsection{Summary of Model-dependency}
	
	It is the nature of the galaxy bias that plays a determining role in how the SC methods are IA model dependent or not. 

The SC2008 in the case of linear bias is simply IA model independent regardless of the IA model. 

However, for non-linear bias, SC2008 has a term in its scaling relation that depends on the ratio $B_{I\delta\delta}/P_{I\delta}$, where $B_{I\delta\delta}$ is the bi-spectrum. 

Therefore, when using the SC2008 scaling relation in situation where non-linear bias and non-linear IA models may be at play, one should expect some loss of efficiency that needs to be estimated from theory due to neglecting the ratio term.  

SC2017 is based on a relation that requires a linear bias model as well as a linear IA model and is thus model dependent. 

\end{document}